\documentclass[%
reprint,
nofootinbib,
 amsmath,amssymb,
 aps,
]{revtex4-2}

\usepackage{graphicx}
\usepackage{dcolumn}
\usepackage{bm}

\newcommand{\s}{\:\!}
\newcommand{\m}{\;\!}
\newcommand{\ket}{\rangle}
\newcommand{\bra}{\langle}

\newcommand{\1}{\mbox{1}\hspace{-0.25em}\mbox{l}}

\begin{document}


\title{Quantization of the damped harmonic oscillator based on a modified Bateman Lagrangian} 

\author{Shinichi Deguchi}
\email[E-mail: ]{deguchi@phys.cst.nihon-u.ac.jp}
\affiliation{Institute of Quantum Science, College of Science and Technology, 
Nihon University, Chiyoda-ku, Tokyo 101-8308, Japan}

\author{Yuki Fujiwara}
\email[E-mail: ]{yfujiwara@phys.cst.nihon-u.ac.jp}
\affiliation{Department of Quantum Science and Technology, Graduate School of Science and Technology, 
Nihon University, Chiyoda-ku, Tokyo 101-8308, Japan}

\date{\today}

\begin{abstract}
An approach to quantization of the damped harmonic oscillator (DHO) is developed 
on the basis of a modified Bateman Lagrangian (MBL);  
thereby some quantum mechanical aspects of the DHO are clarified. 
We treat the energy operator for the DHO, in addition to  
the Hamiltonian operator that is determined from the MBL and corresponds to the total energy of the system. 
It is demonstrated that the energy eigenvalues of the DHO exponentially decrease with time 
and that transitions between the energy eigenstates occur 
in accordance with the Schr\"{o}dinger equation.   
Also, it is pointed out that a new critical parameter discriminates different behaviours of transition probabilities. 
\end{abstract}
%

\maketitle

\section{Introduction} 
Lagrangian-Hamiltonian mechanics of the damped harmonic oscillator (DHO) and  
its applications to quantization of the DHO have been investigated for a long time 
by an enormous number of authors
\cite{Bateman, MorFes, Dekker, Razavy, FesTik, CRV, SVW, BlaJiz, ChrJur, Chruscinski, BanMuk, MajSuz, GLAC, PNC, DegFuj, 
Caldirola, Kanai, Kerner, Hasse, Choi, BFG}. 
One of the most argued Lagrangians of the DHO is the Bateman Lagrangian \cite{Bateman} 
\begin{align}
L_{\rm B}=m\dot{x}\dot{y}+\frac{\gamma}{2}(x\dot{y}-\dot{x}y)-kxy \s. 
\label{1}
\end{align}
This Lagrangian yields the equation of motion of the DHO, 
${m\ddot{x}+\gamma \dot{x} +kx=0}$, 
and has the tractable property that it does not explicitly depend on time. 
However, $L_{\rm B}$ also yields the equation of motion of the amplified harmonic oscillator (AHO), 
${m\ddot{y}-\gamma \dot{y} +ky=0}$. It thus turns out that 
$L_{\rm B}$, in actuality, describes a doubled system consisting of the uncoupled DHO and AHO,  
not the DHO itself. 
The quantization of this system  
has been studied until recently with various interesting ideas 
\cite{FesTik, CRV, SVW, BlaJiz, ChrJur, Chruscinski, BanMuk, MajSuz, GLAC, PNC, DegFuj}.  
However, in the quantization procedure, 
${(x \pm y)/\sqrt{2}}$, rather than $x$ and $y$, are treated as fundamental variables,  
and therefore   
it is quite doubtful whether the DHO itself is correctly quantized in this approach.

In this paper, we develop a novel approach to quantization of the DHO 
to correctly understand the DHO at the quantum level.   
To this end, we propose a {\em modified} Bateman Lagrangian (MBL) 
in order to consistently treat only the DHO. 
We first study the Lagrangian-Hamiltonian mechanics based on the MBL and 
subsequently perform canonical quantization of the DHO  
by utilizing the Lagrangian-Hamiltonian mechanics studied. 
Unlike earlier approaches, we consider the (nonconserved) energy operator for the DHO, 
in addition to the (conserved) Hamiltonian operator that is found from the MBL and 
corresponds to the total energy of the system. 
We show that the energy eigenvalues of the DHO are real and exponentially decrease with time,  
just like the classical energy of the DHO.    
We also show that with the decrease of energy eigenvalues,   
transitions between the energy eigenstates occur in accordance with the Schr\"{o}dinger equation.   
In addition, 
we point out that a new critical parameter discriminates different behaviours of transition probabilities.

\section{Lagrangian-Hamiltonian mechanics based on a MBL} 
Let us begin with the MBL constructed as follows: 
\begin{align}
L_{\rm MB}=L_{\rm B}  
-\frac{1}{2} (\rho \dot{\sigma} -\dot{\rho} \sigma) -\frac{\gamma}{2m} \rho\sigma 
+\lambda( \rho x -\sigma y) \s , 
\label{2}
\end{align}
where $\rho$, $\sigma$, and $\lambda$ are additional real dynamical variables. 
Note that this Lagrangian does not explicitly depend on time. 
From the 5 Euler-Lagrange equations implied by $L_{\rm MB}$, one of which is 
${\rho x=\sigma y}$, 
we can obtain ${\lambda=0}$, 
${2m\dot{\rho} -\gamma\rho=0}$, and ${2m\dot{\sigma} +\gamma\sigma=0}$, in addition to 
the above-mentioned equations of motion for $x$ and $y$ (see Appendix). 
The condition ${\rho x=\sigma y}$, together with ${\rho\sigma >0}$ imposed later under (\ref{3}),  
leads to the fact that the initial phases of $x$ and $y$ are equal modulo $2n\pi$ (${n\in \Bbb{Z}}$) 
(see Appendix). 
We thus see that only one oscillation term exists in this system.

Now we have the canonical coordinates 
${(x, y, \rho, \sigma, \lambda)}$ and their conjugate momenta 
${(p_{x}, p_{y}, p_{\rho}, p_{\sigma}, p_{\lambda})}$ defined from $L_{\rm MB}$. 
Following the Dirac algorithm for constrained systems \cite{Dirac, HRT, HenTei}, we obtain 6 constraints for the 10 canonical variables. 
Hence we actually have 4 independent canonical variables. 
Among several choices, we now choose $(x, p_{x}, \rho, \sigma)$ as independent variables to describe the DHO. 
Accordingly, we have the Hamiltonian that is written in terms of the 4 variables  
${X:=\sqrt{2}\s x}$, ${P:=\sqrt{2}\s p_{x}}$, ${\theta:=(1/2)\ln (\rho/ \sigma)}$, and ${N:=\rho\sigma}$ as
\begin{align}
H =\frac{1}{2m} e^{-2\theta} {P}^{2}
+\frac{1}{2} m{\omega}_{-}^{2} e^{2\theta} {X}^{2} +\frac{\gamma}{2m} N \s, 
\label{3}
\end{align}
where ${\omega_{-} :=\sqrt{ \omega^2-\gamma^2/4m^2 }}$ with ${\omega:=\sqrt{k/m}}$. 
We assume that $\theta$ is real and $N$ is positive real so that $H$ can be positive definite. 
(An inverse Legendre transformation of $H$ leads to a Lagrangian expressed in terms of   
$(X, \theta, \dot{X}, \dot{\theta})$.) 
The non-vanishing Dirac brackets are derived as follows: 
${\{ X, P \}_{\rm D}=1}$, ${\{X, N \}_{\rm D}=-X}$, ${\{P, N \}_{\rm D}=P}$, ${\{\theta, N \}_{\rm D}=1}$.    
Unlike the Caldirola-Kanai Hamiltonian \cite{Caldirola,Kanai}, $H$ does not explicitly depend on time.     
For this reason, $H$ turns out to be a conserved quantity.  
The Hamiltonian $H$ is recognized as the total energy of the system.

The mechanical energy of the DHO is given by  
$E=(m/2)\dot{X}{}^{2} +(m\omega^{2}/2) X^{2}$, 
which can be expressed as 
\begin{align}
E=\frac{1}{2m} \! \left( e^{-2\theta} P -\frac{\gamma}{2} X \right)^{2} +\dfrac{1}{2} m\omega^{2} X^{2} 
\label{4}
\end{align}
by using $\dot{X}=\{ X, H \}_{\rm D}$. 
The conserved Hamiltonian $H$ can be decomposed as ${H=E+Q}$, with $Q$ being identified as  
the heat energy generated in the system.\footnote{With ${X:=\sqrt{2}\s x}$, the equation of motion for $x$ 
is written as ${m\ddot{X}+\gamma \dot{X} +kX=0}$, whose energy integral reads  
$(m/2) \dot{X}^{2} +(m\omega^{2} /2) X^{2} 
+{\gamma \int \dot{X}^{2} dt}=\mbox{constant}$.  
We thus see that ${\gamma \int \dot{X}^{2} dt}$ represents the heat energy generated during
the damped oscillation. Substituting the general solution of the equation of motion for $X$ into 
(\ref{3}), (\ref{4}), and ${\gamma \int \dot{X}^{2} dt}$, we can confirm that 
${Q\s(=H-E)=\gamma \int \dot{X}^{2} dt}$ under the condition ${\theta(0)=0}$. 
Also, under this condition, we can obtain ${H=E=H_{0}}$ from (\ref{3}) and (\ref{4}) when $\gamma=0$. 
Here, $H_{0}$ denotes the Hamiltonian of the ordinary simple harmonic oscillator.}

\section{Canonical quantization} 
Next we perform the canonical quantization of the DHO by replacing $X$, $P$, $\theta$, and $N$ 
with their corresponding Hermitian operators $\hat{X}$, $\hat{P}$, $\hat{\theta}$, and $\hat{N}$, respectively, 
and by setting the commutation relations in accordance with    
${[\hat{A}, \hat{B}\s]=i\hbar \{A, B\}_{\rm D} \1}$. Here, $\1$ denotes the identity operator. 
Through this quantization procedure, we define the Hamiltonian operator $\hat{H}$ and 
the energy operator $\hat{E}$ using (\ref{3}) and (\ref{4}). 
We can verify that ${[\hat{H}, \hat{E}\s]\neq 0}$; hence, $\hat{E}$ is not a conserved quantity as expected.  
The Heisenberg equations 
${i\hbar d\hat{\theta}/dt=[\s\hat{\theta},  \hat{H}\s ]}$ and ${i\hbar d\hat{N}/dt=[\hat{N},  \hat{H}\s ]}$ 
can be solved to yield 
${\hat{\theta}(t)= (\gamma/2m)t \1 +\hat{\theta}_{0}}$ and ${\hat{N}(t)=\hat{N}_{0}}$.  
Here, $\hat{\theta}_{0}$ and $\hat{N}_{0}$ are time-independent operators satisfying  
${[\s\hat{\theta}_{0},  \hat{N}_{0}]=i\hbar \1}$.

We now define the operator 
\begin{align}
\hat{a} 
=\sqrt{\frac{m\omega_{+}}{2\hbar}}\s \varLambda^{\ast} e^{\hat{\theta}} \hat{X}
+i \sqrt{\frac{1}{2\hbar m\omega_{+}}} \s \varLambda\s e^{-\hat{\theta}} \hat{P} \s ,
\label{5}
\end{align}
where 
${\varLambda:=\sqrt{(1+\omega_{+}/\omega)/2} +i\sqrt{(-1+\omega_{+}/\omega)/2}}$ 
with ${\omega_{+} :=\sqrt{ \omega^2 +\gamma^2/4m^2 }}$. 
It is easy to show that 
${[\hat{a}, \hat{a}{}^{\dagger}]=\1}$ and  
${[\hat{a}, \hat{\theta}_{0}]=[\hat{a}{}^{\dagger}, \hat{\theta}_{0}]=0}$.  
In terms of $\hat{a}$, $\hat{a}{}^{\dagger}$, and 
$\hat{N}{}^{\prime}:=\hat{N}+(\hat{X}\hat{P}+\hat{P}\hat{X})/2$, the operator $\hat{H}$ is written as 
\begin{align}
\hat{H}& =\frac{\hbar\omega_{-}^{2}}{\omega} \! 
\left(\hat{a}{}^{\dagger}\hat{a} +\frac{1}{2} \1\right) 
-\frac{\hbar \gamma^{2}}{8m^{2} \omega_{+}}
\left\{ \left( 1 -\frac{i\gamma}{2m\omega} \right) \! \hat{a}{}^{2} \right.  
\notag
\\
& \s\quad \left. +\left( 1 +\frac{i\gamma}{2m\omega} \right) \! \hat{a}{}^{\dagger\s 2} \right\} 
+\frac{\gamma}{2m} \hat{N}^{\prime} .  
\label{6} 
\end{align}
It should be emphasized here that the non-vanishing commutation relations 
for $(\hat{X}, \hat{P}, \hat{\theta}, \hat{N}^{\prime})$ are only 
$[\hat{X}, \hat{P}]=i\hbar\1$ and ${[\s\hat{\theta}, \hat{N}^{\prime}]=i\hbar\1}$. 
We thus see that the canonical conjugate operator to $\hat{\theta}$ is $\hat{N}^{\prime}$ rather than $\hat{N}$. 
The energy operator can be written as  
\begin{align}
\hat{E}=\hbar \omega e^{-2\hat{\theta}(t)}
\!\left( \hat{a}^{\dagger} \hat{a} +\frac{1}{2} \1\right) 
\label{7}
\end{align}
with $\hat{\theta}(t)=(\gamma/2m)t \1 +\hat{\theta}_{0}$.

Now we introduce the ground state vector ${|0, t\ket}$ specified by 
${\hat{a}(t){|0, t\ket}=0}$ and ${\hat{\theta}_{0} {|0, t\ket}=0}$. 
The second condition is necessary to reproduce the simple harmonic oscillator system when $\gamma=0$.  
The Fock basis vectors are constructed as  
${|n, t \ket =(1/{\sqrt{n!}}) \big(\hat{a}{}^{\dagger}(t) \big){}^{n} |0, t\ket}$ $(n=0, 1, 2, \ldots)$,  
which obviously satisfy ${\hat{\theta}_{0} |n, t\ket=0}$. The energy eigenvalue equation is found to be 
$\hat{E} |n, t \ket =E_{n} |n, t \ket$ with the energy eigenvalues 
\begin{align}
{E}_{n}=\hbar \s\omega e^{-\gamma t/m} \!\left( n+\frac{1}{2} \right)  . 
\label{8}
\end{align}
All the energy eigenvalues decrease exponentially with time and eventually vanish 
in the limit ${t \rightarrow \infty}$, 
while maintaining the energy distribution with equal intervals at each time point $t$. 
Incidentally, the classical energy of the DHO is also proportional to $e^{-\gamma t/m}$.   
To the best of our knowledge, 
(\ref{8}) has not been found in the earlier literature on quantization of the DHO.\footnote{A similar but different 
expression,  ${\hbar (\omega^2 / \omega_{-}) \s e^{-\gamma t/m} (n+1/2)}$,  
has been derived as an energy expectation value of the DHO \cite{Kerner,Hasse,Choi,BFG}. 
This expression, however, behaves in a strange manner such that 
it diverges in the critical damping limit $\omega_{-}  \rightarrow 0$. }

\section{The Schr\"{o}dinger picture} 
The time-evolution operator is given by 
${\hat{U}=\exp (-i\hat{H}t/\hbar)}$. 
Here, $\hat{H}$ is understood as $\hat{H}(0)$ because $\hat{H}$ is a conserved quantity.  
We define the time-independent operators $\hat{X}_{\rm S}$ and $\hat{P}_{\rm S}$ in the Schr\"{o}dinger picture 
by ${\hat{X}_{\rm S}=\hat{U} \hat{X} \hat{U}{}^{\dagger}}$ and 
${\hat{P}_{\rm S}=\hat{U} \hat{P} \hat{U}{}^{\dagger}}$ \cite{Sakurai}. 
Similarly, we define ${|n, t\ket_{\rm S}=\hat{U} |n, t\ket}$. 
In terms of $|n, t\ket_{\rm S}$, the condition ${\hat{\theta}_{0} |n, t\ket=0}$ reads 
\begin{align}
\hat{\theta}_{0} |n, t\ket_{\rm S}=\frac{\gamma}{2m} t \s |n, t\ket_{\rm S} \s.
\label{9}
\end{align}
Equation (\ref{9}) implies that in the Schr\"{o}dinger picture, 
$(2m/\gamma)\hat{\theta}_{0}$ behaves as a time operator.  
This operator is well-defined, because the canonical conjugate operator   
$\hat{N}^{\prime}_{0}(:=\hat{N}^{\prime}(0))$ 
can possess eigenvalues unbounded below and above, 
unlike $\hat{N}_{0}$ whose eigenvalues are assumed to be positive so that 
the condition $N_{0}>0$ at the classical level would be inherited. 
Combining (\ref{9}) with ${[\s\hat{\theta}_{0}, \hat{N}_{0}^{\prime}]=i\hbar\1}$ leads to 
\begin{align}
{}_{\rm S} \bra n, t | \hat{N}^{\prime}_{0} 
=-i\hbar \frac{2m}{\gamma} \frac{d}{dt} \s {}_{\rm S} \bra n, t | \,. 
\label{10}
\end{align}

Using (\ref{9}), we can show that 
$\hat{a}(0)|n, t\ket_{\rm S} =\hat{a}_{\rm S}(t)|n, t\ket_{\rm S}$, 
$\hat{a}{}^{\dagger}(0)|n, t\ket_{\rm S} =\hat{a}{}^{\dagger}_{\rm S}(t)|n, t\ket_{\rm S}$,  
and 
furthermore 
$|n, t \ket_{\rm S} = (1/{\sqrt{n!}}) \big(\hat{a}_{\rm S}^{\dagger}(t) \big){}^{n} |0, t\ket_{\rm S}$
with $\hat{a}_{\rm S}(t) |0, t\ket_{\rm S}=0$, where  
\begin{align}
\hat{a}_{\rm S}(t) 
=\sqrt{\frac{m\omega_{+}}{2\hbar}}\s \varLambda^{\ast} e^{\gamma t/2m} \hat{X}_{\rm S}
+i \sqrt{\frac{1}{2\hbar m\omega_{+}}} \s \varLambda\s e^{-\gamma t/2m} \hat{P}_{\rm S} \s . 
\label{11}
\end{align}
The energy eigenfunction corresponding to the energy eigenvalue $E_{n}$ is derived as follows: 
\begin{align}
& \phi_{n}(X, t) :=\bra X |n, t \ket_{\rm S} 
\notag
\\
&=\frac{1}{\sqrt{2^{n} n!}} \left(\frac{m\omega}{\pi\hbar}\right)^{\! 1/4} 
\left(\frac{\omega-\frac{i\gamma}{{2m}}}{\omega+\frac{i\gamma}{{2m}}}\right)^{\! n/4} \!
H_{n} \! \left(\sqrt{\frac{m\omega}{\hbar}}\m e^{\gamma t/2m} X \right) 
\notag
\\
&\,\quad \times \exp \! \left[\s \frac{\gamma}{4m} t -\frac{m}{2\hbar} \! \left( \omega -\frac{i\gamma}{2m} \right) 
\! e^{\gamma t/m} X^2 \s\right] , 
\label{12}
\end{align}
where $H_{n}$ denotes the $n$th Hermite polynomial. 
It is easy to verify that $\int \phi_{n}^{\ast}(X, t) \phi_{n^{\prime}}(X, t) dX=\delta_{nn^{\prime}}$.  
In FIG. 1,  we show the graphs of $|\phi_{n}|^{2}$ ${(n=0,1,2)}$ plotted as functions of $X$ 
at ${t=0}$ and ${t=250}$ 
for the fixed values ${m=10}$, ${\omega=1}$, ${\gamma=0.1}$, and ${\hbar=1}$. 
As ${t \rightarrow \infty}$, 
$|\phi_{n}|^{2}$ infinitely increases in an infinitesimal neighborhood, $\frak{N}$, of the origin ${X=0}$, 
while decreasing to zero in the domain ${\Bbb{R} \setminus \frak{N}}$. 
When ${\gamma=0}$, $\phi_{n}$ reduces to the $n$th energy eigenfunction of the ordinary simple harmonic oscillator. 
%
\begin{figure}
 \begin{center}
   \includegraphics*[width = 7.7cm]{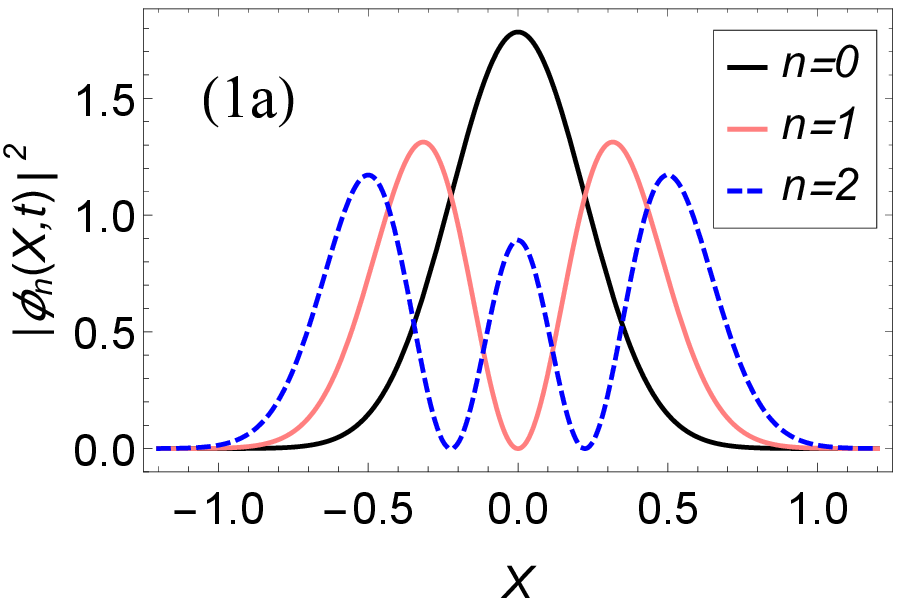}
\end{center}
\begin{center}
   \includegraphics*[width = 7.7cm]{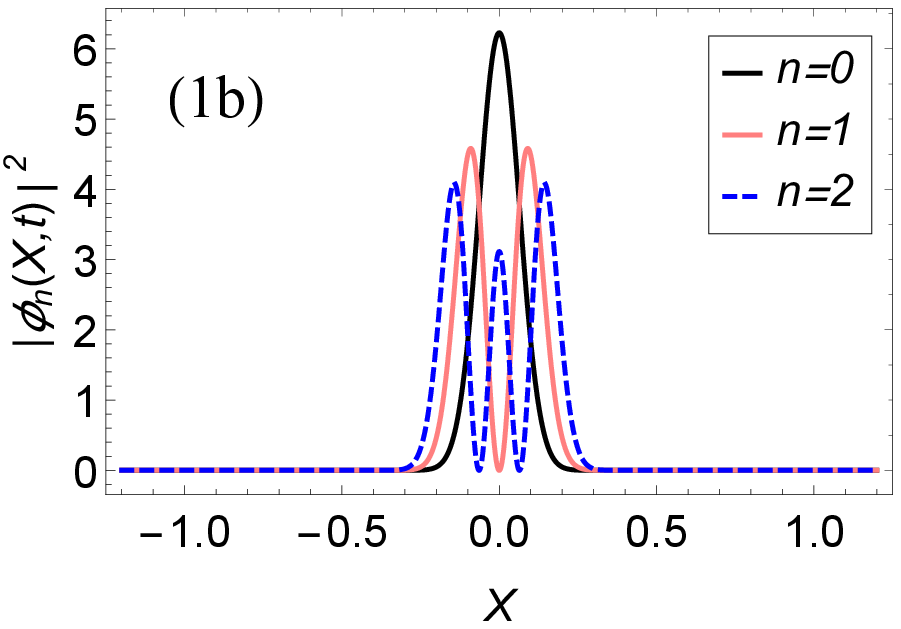}
   \caption{Figures (1a) and (1b) show the graphs of $|\phi_{n}(X,t)|^{2}$ ${(n=0,1,2)}$ at ${t=0}$ and ${t=250}$, 
   respectively.}\label{fig_1}
 \end{center}
\end{figure}
%

\section{The Schr\"{o}dinger equation and its solutions} 
Let $| \psi (t)\ket$ be a state vector that 
can be expanded over the Fock basis ${\{\m|n, t \ket_{\rm S} \}}$.  
Then the Schr\"{o}dinger equation for the present system,  
${i\hbar d \s|\psi(t)\ket /dt=\hat{H}(0) |\psi(t)\ket }$, can be written as 
\begin{align}
i\hbar \frac{d}{dt} | \psi(t) \rangle =\hat{H}_{\rm S}(t) |\psi(t) \rangle \s,    
\label{13}
\end{align}
where $\hat{H}_{\rm S}(t)$ is defined by replacing $\hat{a}$, $\hat{a}{}^{\dagger}$, and $\hat{N}^{\prime}$ 
in (\ref{6}) with $\hat{a}_{\rm S}(t)$, $\hat{a}_{\rm S}^{\dagger}(t)$, and $\hat{N}_{0}^{\prime}$, respectively. 
Now we expand $|\psi(t) \ket$ as 
${|\psi(t) \rangle =\sum_{n} c_{n} (t) 
\exp \! \big[ (i/\hbar) \int_{0}^{t} \varTheta_{n}(t^{\prime}) dt^{\prime} \big] |n, t \ket_{\rm S}}$   
with 
${\varTheta_{n}(t):={}_{\rm S} \bra n, t | 
\big[^{\s} i\hbar d/dt -\hat{H}_{\rm S}(t) \big] |n, t \ket_{\rm S}}$  
\cite{OmoKam, KamOmo}. 
Here the normalization condition ${\sum_{n} |c_{n}(t)|^{2} =1}$ is understood. 
Substituting this $|\psi(t)\ket$ into (\ref{13}), we obtain  
\begin{align}
\frac{dc_{n}(t)}{dt} &=\sum_{n^{\prime} (\neq n)} 
\frac{1}{E_{n}-E_{n^{\prime}}} 
\m {}_{\rm S} \bra n, t | \frac{D\hat{E}_{\rm{S}}(t)}{Dt} | n^{\prime}, t \s\ket_{\rm S} 
\notag
\\
& \s\quad \times c_{n^{\prime}}(t) \exp \!\left[ \s \frac{i}{\hbar} \int_{0}^{t} 
\left\{ \varTheta_{n^{\prime}}(t^{\prime}) - \varTheta_{n}(t^{\prime}) \right\} dt^{\prime} \right] 
\label{14}
\end{align} 
with   
${D\hat{E}_{\rm{S}}(t)/Dt :=d\hat{E}_{\rm{S}}(t)/dt +(1/i\hbar) 
[ \hat{E}_{\rm S}(t), \hat{H}_{\rm S}(t) ]^{\s}}$,   
where 
$\hat{E}_{\rm S}(t)$ is defined by replacing $\hat{a}$, $\hat{a}{}^{\dagger}$, and $\hat{\theta}(t)$ 
in (\ref{7}) with $\hat{a}_{\rm S}(t)$, $\hat{a}_{\rm S}^{\dagger}(t)$, and $\hat{\theta}_{0}$, respectively.  
In deriving (\ref{14}), ${\hat{E}_{\rm S}(t) |n, t \ket_{\rm S} =E_{n} |n, t \ket_{\rm S}}$ has been used. 
It is remarkable that in $\varTheta_{n}(t)$, the geometric phase  
${{}_{\rm S} \bra n, t | \s i\hbar d/dt |n, t \ket_{\rm S}}$ 
is cancelled out with 
${ {}_{\rm S} \bra n, t |(\gamma/2m) \hat{N}_{0}^{\prime} |n, t \ket_{\rm S}}$ 
by means of (\ref{10}).  Consequently, $\varTheta_{n}(t)$ is conveniently simplified and reduces to 
${\varTheta_{n}=-\hbar(\omega_{-}^{2}/\omega) (n+1/2)}$. 
The wave function is then found to be 
\begin{align}
\psi(X, t) :=\bra X |\psi(t) \rangle =\sum_{n} c_{n}(t) 
e^{(i/\hbar) \varTheta_{n} t} \phi_{n}(X,t) \s. 
\label{15}
\end{align}
We see from (\ref{12}) that the dispersion of the probability density $|\psi(X, t)|^{2}$ 
decreases with time and ultimately becomes zero, maintaining 
$\int |\psi(X, t)|^{2} dX=1$.  
This result is consistent with the classical motion of the DHO.

After some calculation, (\ref{14}) becomes 
\begin{align}
\frac{dc_{n}(t)}{dt} &=\frac{\gamma}{4m} 
\left\{ -\sqrt{(n+1)(n+2)} \, e^{-i(2\alpha t+\beta)} c_{n+2}(t) 
\right. 
\notag
\\
& 
\s \quad +\left. 
\sqrt{n(n-1)} \, e^{i(2\alpha t+\beta)} c_{n-2}(t) \right\} ,  
\label{16} 
\end{align}
where $\alpha$ and $\beta$ are defined by ${\alpha=\omega_{-}^{2}/\omega}$ and 
$e^{i\beta}=(\omega +i\gamma/2m)/\omega_{+}$, respectively.  
We here impose the initial condition $c_{n}(0)=\delta_{nl}$ $(l=0, 1, 2, \ldots)$ 
so that the initial state would be $|^{\s} l, 0 \ket_{\rm S}$  
and $\psi(X, 0)=\phi_{l} (X, 0)$ can hold accordingly. 
The solutions of the differential-difference equation (\ref{16}) can be obtained by solving the 
partial differential equation 
\begin{align}
\frac{\partial G}{\partial t}=-\left\{
\frac{\gamma}{4m} \left(\frac{\partial^{2}}{\partial q^{2}} -q^{2} \right) 
+i \alpha q \frac{\partial}{\partial q} \right\} G 
\label{17}
\end{align}
for ${G(q, t):=\sum_{n} q^{n} e^{-in(\alpha t+\beta/2)} c_{n}(t) /\sqrt{n!}}$
under the conditions 
${G(q, 0)=q^{l} e^{-il\beta/2} / \sqrt{l\s !}}$ and ${G(0,t)=c_{0}(t)}$. 
(As for analytically solving differential-difference equations, see, e.g., 
Refs. \cite{Razavy, Razavy_2, ISSSS}.)  
In the following, we investigate the cases $l=0$ and $l=2$ in particular, although 
the other cases can be explicated.

\subsection{Case ${l=0}$}
In the case ${l=0}$, the initial state is the ground state specified by $|0, 0 \ket_{\rm S}$. 
The solution of (\ref{17}) is then found to be 
\begin{align}
G_{0} (q, t)&=\sqrt{\xi} \s e^{i\alpha t/2} \! \left\{\cosh \! \left( \zeta+\xi \gamma t/2m \right) \right\}\!{}^{-1/2} 
\notag 
\\
& \quad \s \times \exp \! 
\left[ \frac{\sinh \!\big(\xi \gamma t/2m \big)}{2\cosh \!\big( \zeta +\xi \gamma t/2m \big)}\m q^{2} \right] , 
\label{18}
\end{align} 
where $\xi$ and $\zeta$ are defined by 
${\xi=\big(1-4m^{2} \alpha^{2}/\gamma^{2} \big){}^{1/2}}$ 
and ${e^{\pm\zeta} =\xi \pm 2im\alpha/\gamma}$, respectively.  
It is easily verified that ${G_{0}(q, 0)=1}$.  
The solution of (\ref{16}) can be derived from (\ref{18}) as follows: 
\begin{align} 
c_{n,0} (t) &=\frac{1}{\sqrt{n!}} e^{in(\alpha t+\beta/2)} 
\frac{\partial^{n}}{\partial q^{n}} G_{0} (q,t) \bigg|_{q=0} 
\notag
\\
&=\left\{ \begin{aligned} 
\;\: \frac{(n-1)!!}{\sqrt{n!}} \sqrt{\xi} \s e^{i(n+1/2)\alpha t} e^{in\beta/2} \qquad 
\notag
\\
\; \s\times \m \frac
{\left\{ \sinh \!\big(\xi \gamma t/2m \big) \right\}{}^{\! n/2} }
{\left\{ \cosh \!\big( \zeta+\xi \gamma t/2m \big) \right\}{}^{\!(n+1)/2}} \quad\,\; 
\notag 
\\[1pt]
\mbox {for $\, n=0,2,4, \ldots,$} 
\qquad \qquad \quad 
\\[3pt]
0 \quad \!\s
\mbox {for $\, n=1,3,5, \ldots,$} 
\qquad \qquad \quad 
\end{aligned} \right. 
\tag{19}
\label{19}
\end{align}
which certainly satisfies the conditions ${\sum_{n} |c_{n,0}(t)|^{2} =1}$, 
${c_{n,0}(0)=\delta_{n0}}\s$, and ${G_{0}(0,t)=c_{0,0}(t)}$.

Now we evaluate the transition probability from $|0, 0 \ket_{\rm S}$ to $|n, t \ket_{\rm S}$,   
described by $|c_{n,0} (t)|^{2}$. 
Since no transition occurs when $n$ is odd, we hereafter consider only the cases in which $n$ is even. 
As seen from (\ref{19}), the time evolution of $|c_{n,0} (t)|^{2}$ essentially depends on 
$e^{\pm\xi \gamma t/2m}$. For this reason, 
it is necessary to separately evaluate $|c_{n,0} (t)|^{2}$ in the following three cases:   
(a)~${(0\le\s) \: \gamma <\gamma_{\ast}}$, (b)~${\gamma =\gamma_{\ast}}$, and (c)~${(2m\omega>)\: \gamma >\gamma_{\ast}}$.
Here, $\gamma_{\ast}$ stands for the critical constant parameter $(\sqrt{5}-1) m\omega \simeq 1.236 m\omega$, and 
${2m\omega >\gamma}$ is the classical condition for the damped oscillation.

In the case (a), $\xi$ becomes a purely imaginary number, 
and accordingly $|c_{n,0} (t)|^{2}$ becomes a periodic function. 
In Fig. (2a), we show the graphs of $|c_{n,0} (t)|^{2}$ $(n=0,2,4,6)$ for the fixed values 
${m=\omega=1}$ and ${\gamma=1}$, which satisfy ${\gamma < \gamma_{\ast}}$.  
The transition probabilities $|c_{n,0} (t)|^{2}$ change periodically with the same period.

In the case (b), $\xi$ vanishes, and hence we need to expand Eq. (\ref{19}) around ${\xi=0}$ 
to obtain  
\begin{align}
c_{n,0} (t) 
=\frac{(n-1)!!}{\sqrt{n!}} \s e^{i(n+1/2)\alpha t} e^{in\beta/2} 
\frac{(\gamma t/2m){}^{n/2}}
{(1+i\alpha t){}^{(n+1)/2}} \,. 
\tag{20}
\end{align}
Clearly, $|c_{n,0}(t)|^{2}$ is an irrational function. 
Figure (2b) shows the graphs of $|c_{n,0} (t)|^{2}$ ${(n=0,2,4,6)}$ for the fixed values 
${m=\omega=1}$ and ${\gamma=\sqrt{5}-1}$, which satisfy ${\gamma = \gamma_{\ast}}$.  
The transition probability $|c_{0,0} (t)|^{2}$ decreases monotonically, while 
$|c_{n,0} (t)|^{2}$ ${(n=2, 4, 6, \ldots)}$ increase once in the order of $n$  
and subsequently decrease monotonically.

In the case (c), $\xi$ becomes a positive real number, and accordingly $|c_{n,0} (t)|^{2}$ becomes 
a combination of real hyperbolic functions.  
Figure (2c) shows the graphs of $|c_{n,0} (t)|^{2}$ ${(n=0,2,4,6)}$ for the fixed values 
${m=\omega=1}$ and ${\gamma=1.5}$, which satisfy ${\gamma> \gamma_{\ast}}$.  
The shapes of the curves in Fig. (2c) are similar to those in Fig. (2b); the differences, 
such as the rates of changes, are essentially due to 
the presence of $e^{\pm\xi \gamma t/2m}$ ($\xi>0$). 
%
\begin{figure}
 \begin{center}
   \includegraphics*[width = 7.7cm]{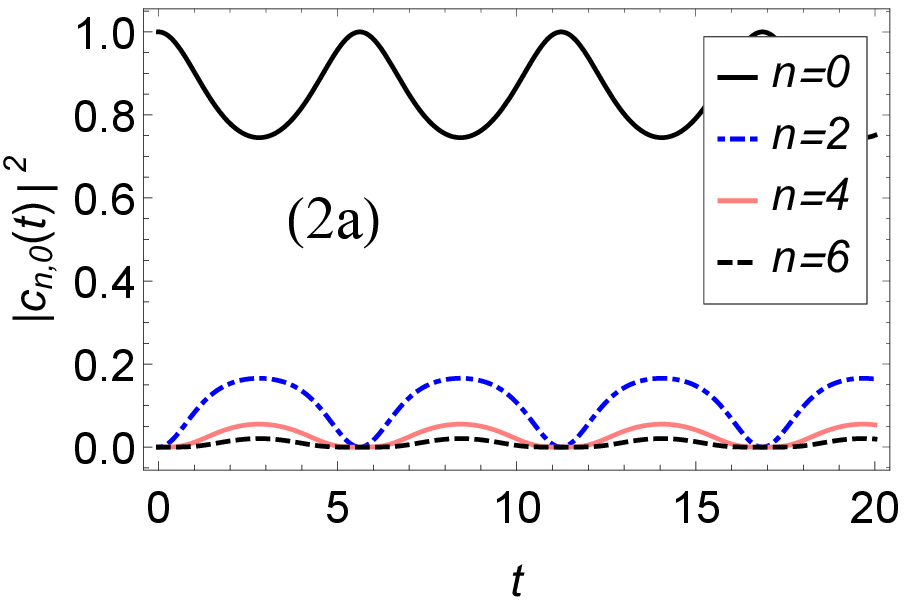}
 \end{center}
 \begin{center}
   \includegraphics*[width = 7.7cm]{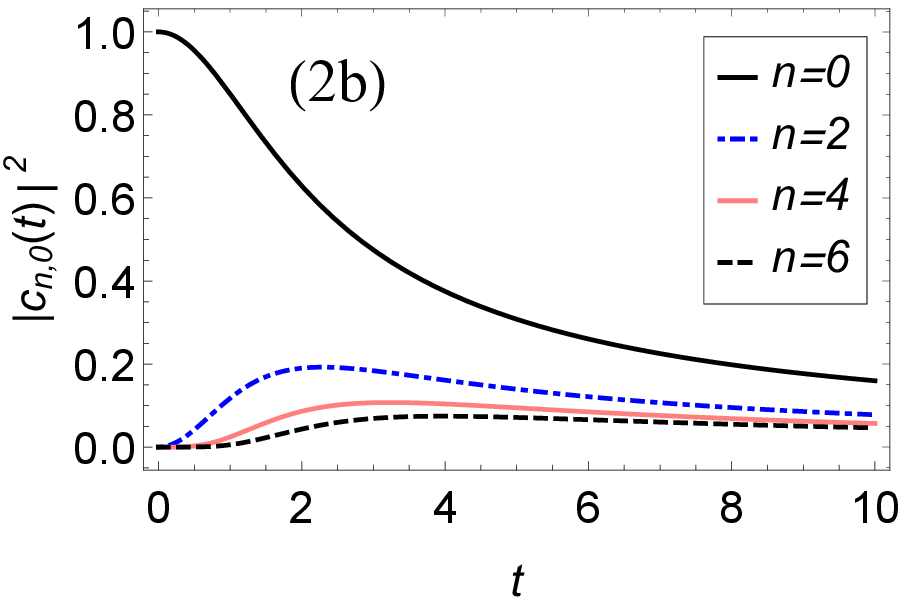}
 \end{center}
 \begin{center}
   \includegraphics*[width = 7.7cm]{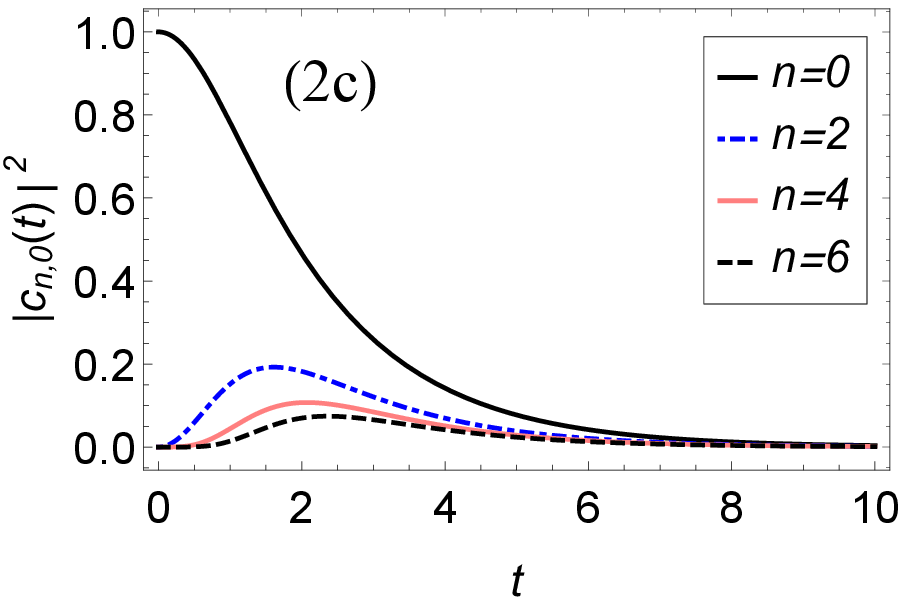}
   \caption{Figures (2a), (2b), and (2c) show the graphs of $|c_{n,0} (t)|^{2}$ ${(n=0,2,4,6)}$ plotted 
   in the cases (a), (b), and (c), respectively.}\label{fig_2} 
 \end{center}
\end{figure}
%

\subsection{Case ${l=2}$} 

In the case ${l=2}$, the initial state is the 2nd excited state specified by $|2, 0 \ket_{\rm S}$.    
We can obtain the solution of (\ref{17}) for ${l=2}$ and denote it as $G_{2} (q, t)$.   
This satisfies the condition ${G_{2}(q, 0)=q^{2} e^{-i\beta}/\sqrt{2}}$. 
The corresponding solution of (\ref{16}) is found to be  
\begin{align} 
c_{n,2} (t) &=\frac{1}{\sqrt{n!}} e^{in(\alpha t+\beta/2)} 
\frac{\partial^{n}}{\partial q^{n}} G_{2} (q,t) \bigg|_{q=0} 
\notag
\\
&=\left\{ \begin{aligned} 
\frac{(n-1)!!}{\sqrt{2n!}} \sqrt{\xi} \s e^{i(n+1/2)\alpha t} e^{i(n/2-1)\beta} \qquad \;\;\, 
\notag
\\
\;\; \times \s \left\{ -\sinh \!\big(\xi \gamma t/2m \big) 
+\frac{n\s\xi^{2}}{ \sinh \!\big(\xi \gamma t/2m \big)} \right\} 
\notag
\\
\; \m\times \, \frac
{\left\{ \sinh \!\big(\xi \gamma t/2m \big) \right\}{}^{\! n/2} }
{\left\{ \cosh \!\big( \zeta+\xi \gamma t/2m \big) \right\}{}^{\!(n+3)/2}} \qquad \quad \;\;\; 
\\[1pt]
\mbox {for $\, n=0,2,4, \ldots,$} 
\qquad \qquad \qquad \quad \;\s
\\[3pt]
0  \quad 
\mbox {for $\, n=1,3,5, \ldots,$} 
\qquad \qquad \qquad \quad \;\s
\end{aligned} \right. 
\tag{21}
\label{21}
\end{align}
which certainly satisfies the conditions ${\sum_{n} |c_{n,2}(t)|^{2} =1}$, 
${c_{n,2}(0)=\delta_{n2}}\s$, and ${G_{2}(0,t)=c_{0,2}(t)}$.

We next evaluate the transition probability from $|2, 0 \ket_{\rm S}$ to $|n, t \ket_{\rm S}$,   
described by $|c_{n,2} (t)|^{2}$. 
As in the case $l=0$, it is sufficient to consider only the cases in which $n$ is even. 
Since the time evolution of $|c_{n,2} (t)|^{2}$ intrinsically depends on $e^{\pm\xi \gamma t/2m}$, 
we need to separately evaluate $|c_{n,2} (t)|^{2}$ in the above mentioned three cases (a), (b), and (c).

In the case (a),  $|c_{n,2} (t)|^{2}$ becomes a periodic function. 
Figure (3a) shows the graphs of $|c_{n,2} (t)|^{2}$ $(n=0,2,4,6)$ for the fixed values 
${m=\omega=1}$ and ${\gamma=1}$.  
It is confirmed that 
the transition probabilities $|c_{n,2} (t)|^{2}$ change periodically with the same period.

In the case (b), $\xi$ vanishes, and it is necessary to expand Eq. (\ref{21}) around ${\xi=0}$ to obtain  
\begin{align}
c_{n,2} (t) 
& =\frac{(n-1)!!}{\sqrt{2n!}} \s e^{i(n+1/2)\alpha t} e^{i(n/2 -1)\beta} 
\notag
\\
& \quad \m \times \left(-\frac{\gamma t}{2m} +\frac{2mn}{\gamma t} \right) 
\frac{(\gamma t/2m){}^{n/2}}
{(1+i\alpha t){}^{(n+3)/2}} \,.   
\tag{22}
\end{align}
Obviously, $|c_{n,2}(t)|^{2}$ is an irrational function. 
Figure (3b) shows the graphs of $|c_{n,2} (t)|^{2}$ ${(n=0,2,4,6)}$ for the fixed values 
${m=\omega=1}$, and ${\gamma=\sqrt{5}-1}$.

In the case (c), $|c_{n,2} (t)|^{2}$ becomes a combination of real hyperbolic functions.  
Figure (3c) shows the graphs of $|c_{n,2} (t)|^{2}$ ${(n=0,2,4,6)}$ for the fixed values 
${m=\omega=1}$ and ${\gamma=1.5}$.  
%
\begin{figure}
 \begin{center}
   \includegraphics*[width = 7.7cm]{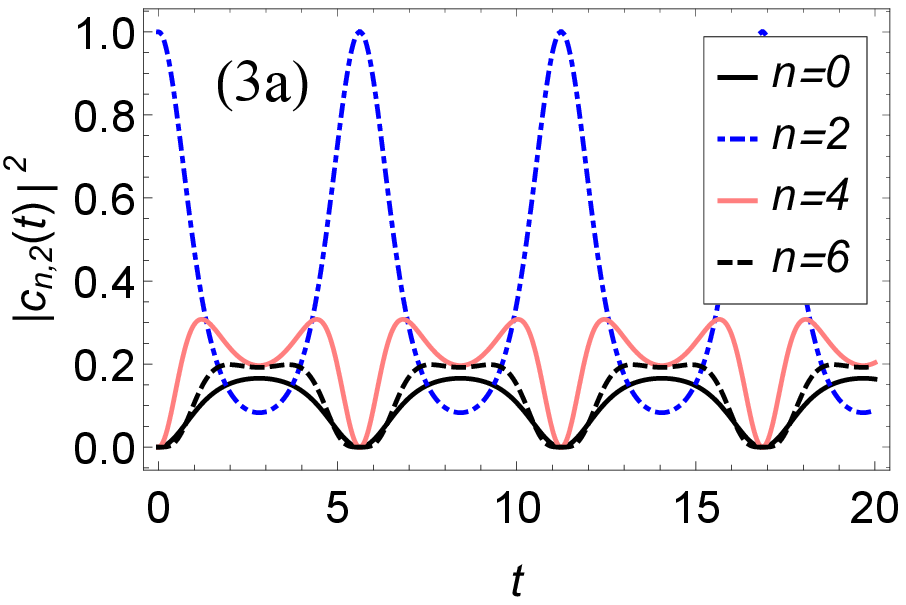}
 \end{center}
 \begin{center}
   \includegraphics*[width = 7.7cm]{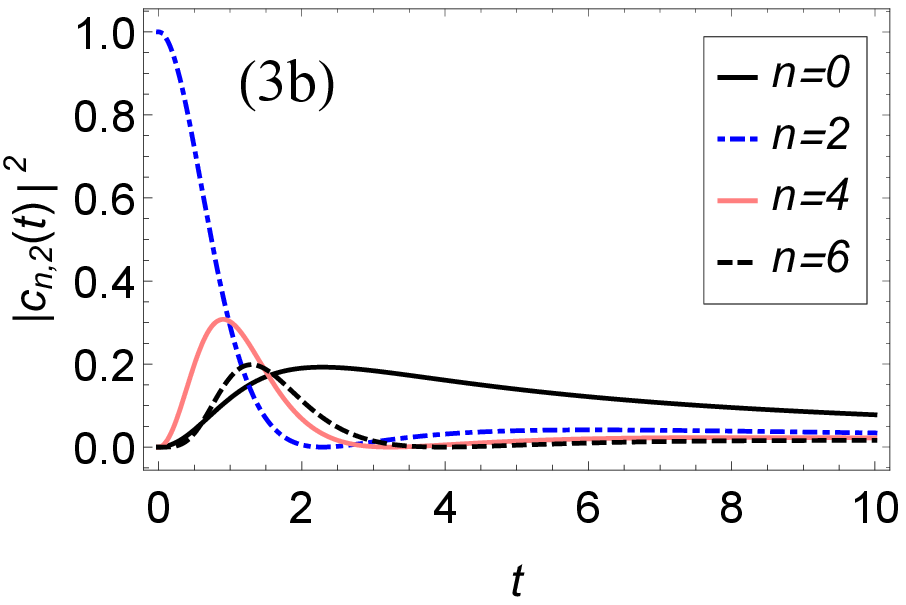}
 \end{center}
 \begin{center}
   \includegraphics*[width = 7.7cm]{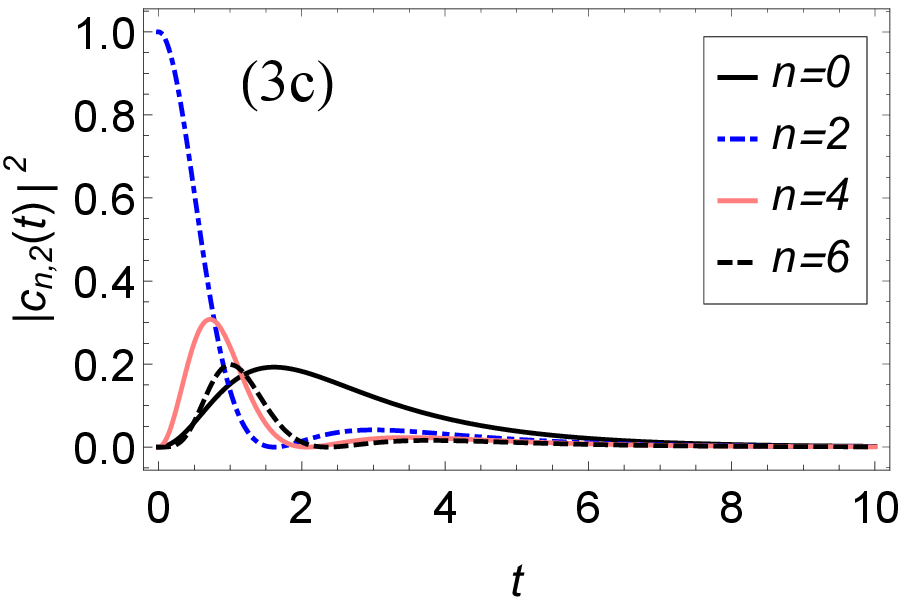}
   \caption{Figures (3a), (3b), and (3c) show the graphs of $|c_{n,2} (t)|^{2}$ ${(n=0,2,4,6)}$ plotted    
   in the cases (a), (b), and (c), respectively.}\label{fig_3}
 \end{center}
\end{figure}
%

Comparing the graphs in FIGs. 2 and 3 plotted for the same $\gamma$ and $n$, we observe that 
most of the graphs in FIG. 3 have more inflection points than the corresponding graphs in FIG. 2. 
Such details on the graphs of $|c_{n,l} (t)|^{2}$ should be examined analytically 
in the case of arbitrary $l$ and $n$.

\section{Concluding remarks} 

In conclusion, the DHO at the quantum level is understood as the one 
whose energy eigenvalues with equal energy intervals decrease exponentially with time 
and that involves transitions between the energy eigenstates in association with the decrease of energy eigenvalues. 
To the best of our knowledge, no such quantum mechanical aspects of the DHO have been illustrated in earlier literature. 
It is remarkable that in addition to the classical critical parameter $2m\omega$, 
the new critical parameter ${\gamma_{\ast} \equiv (\sqrt{5}-1) m\omega}$ appears at the quantum level. 
This parameter discriminates different behaviours of ${|c_{n,l}(t)|^{2}}$ under time evolution.

We first considered the doubled system with the dynamical variables $x$ and $y$. 
The doubling of dynamical variables is a common strategy for dealing with dissipative systems 
such as the DHO \cite{TakUme, Umezawa, Galley, GTS, Polonyi, DNS}, 
regardless of whether or not the additional variables represent the degrees of freedom 
of a heat bath or environment. 
In fact, Galley developed a new framework of Lagrangian-Hamiltonian mechanics for generic dissipative systems 
by means of the doubling of dynamical variables \cite{Galley}. 
In this framework, after all variations are performed, each doubled variables are reduced to a single physical variable 
by imposing the condition called physical limit by hand.   
In our approach, instead, an alternative condition, $\rho x=\sigma y$, is imposed at the Lagrangian level as in (\ref{2}).

In this paper, we have not explicitly treated the degrees of freedom of a heat bath or environment, 
although the heat energy ${Q=H-E}$ has been taken into account. 
In this sense, our approach is, so to speak, phenomenological. 
It would be interesting to generalize our phenomenological approach to other dissipative systems.

\appendix*
\section{}

This Appendix is devoted to deriving the equations mentioned under (\ref{2}) 
and to examining their general solutions.

From $L_{\rm MB}$, we obtain the Euler-Lagrange equations 
\begin{subequations}
\label{A1}
\begin{align}
m\ddot{x}+\gamma \dot{x} +kx +\lambda \sigma &=0 \,,
\label{A1a}
\\
m\ddot{y}-\gamma \dot{y} +ky -\lambda \rho &=0 \,,
\label{A1b}
\\
2m\dot{\rho} -\gamma \rho -2m\lambda y &=0 \,,
\label{A1c}
\\
2m\dot{\sigma} +\gamma\sigma -2m \lambda x &=0 \,,
\label{A1d}
\\
\rho x-\sigma y&=0 \,.
\label{A1e}
\end{align}
\end{subequations}
To avoid the reduction to the original Bateman model, 
we here assume that ${\rho\neq 0}$ and ${\sigma \neq 0}$. 
Then (\ref{A1e}) can be written as $y=(\rho/\sigma) x$. 
Using (\ref{A1c}), (\ref{A1d}), and (\ref{A1e}), we have 
\begin{align}
\dot{y}&=\frac{\rho}{\sigma} \left(\dot{x}+\frac{\gamma}{m} x \right) ,
\label{A2}
\\
\ddot{y}&=\frac{\rho}{\sigma} \left( \ddot{x}+\frac{2\gamma}{m} \dot{x}+\frac{\gamma^2}{m^2} x\right) .
\label{A3}
\end{align}
Substituting ${y=(\rho/\sigma) x}$, (\ref{A2}), and (\ref{A3}) into Eq. (\ref{A1b}) leads to 
\begin{align}
m\ddot{x}+\gamma \dot{x} +kx -\lambda \sigma=0 \,.
\label{A4}
\end{align}
From (\ref{A1a}) and (\ref{A4}), we have  
\begin{align}
m\ddot{x}+\gamma \dot{x} +kx &=0 \,, 
\label{A5}
\\
\lambda &=0 \,, 
\label{A6}
\end{align}
because ${\sigma\neq 0}$. 
In this way, $\lambda$ is automatically determined to be 0;   
as a result, (\ref{A1b}), (\ref{A1c}), and (\ref{A1d}) become 
\begin{subequations}
\label{A7}
\begin{align}
m\ddot{y}-\gamma \dot{y} +ky &=0 \,,
\label{A7a}
\\
2m\dot{\rho} -\gamma \rho &=0 \,,
\label{A7b}
\\
2m\dot{\sigma} +\gamma\sigma &=0 \,,  
\label{A7c}
\end{align}
\end{subequations}
respectively. 
Thus we can naturally derive (\ref{A5})--(\ref{A7}) and (\ref{A1e}), 
namely the equations mentioned under (\ref{2}), 
from $L_{\rm MB}$.

The general solutions of  (\ref{A5}), (\ref{A7a}), (\ref{A7b}), and (\ref{A7c}) 
are, respectively, found to be  
\begin{subequations}
\label{A8}
\begin{align}
x(t) &=x_{0} e^{-\gamma t/2m} \sin(\omega_{-} t +\varphi) \,,
\label{A8a}
\\
y(t) &=y_{0} e^{\gamma t/2m} \sin(\omega_{-} t +\chi) \,,
\label{A8b}
\\
\rho(t) &=\rho_{0} e^{\gamma t/2m} , 
\label{A8c}
\\
\sigma(t) &=\sigma_{0} e^{-\gamma t/2m} , 
\label{A8d}
\end{align}
\end{subequations}
where $x_{0}$ and $y_{0}$ are positive real constants, 
and $\varphi$, $\chi$, $\rho_{0}$, and $\sigma_{0}$ are real constants. 
Substituting (\ref{A8a})--(\ref{A8d}) into (\ref{A1e}) gives   
\begin{align}
\rho_{0} x_{0} \sin(\omega_{-} t +\varphi) =\sigma_{0} y_{0} \sin(\omega_{-} t +\chi) \,. 
\label{A9}
\end{align}
Dividing (\ref{A9}) by its derivative with respect to $t$, we have 
${\tan(\omega_{-} t +\varphi) =\tan(\omega_{-} t +\chi)}$, which implies that 
${\chi=\varphi +n\pi}$ (${n\in \Bbb{Z}}$).  
Substituting this into Eq. (\ref{A9}) yields 
${\rho_{0} x_{0}=(-1)^{n} \sigma_{0} y_{0}}$. 
Since ${\rho_{0} \sigma_{0}=\rho\sigma=N>0}$ is assumed under (\ref{3}), 
in addition to ${x_{0}>}0$ and ${y_{0}>0}$, we conclude that $n$ is even.  
Hence, the initial phases $\varphi$ and $\chi$ are equal modulo $2\pi n$ ($n\in \Bbb{Z}$). 
We thus see that only one oscillation term, ${\sin(\omega_{-} t +\varphi)}$, exists in the present system.


\begin{thebibliography}{00}

\bibitem{Bateman}
H.~Bateman, 
On dissipative systems and related variational principles, 
Phys. Rev. {\bf 38}, 815 (1931) . 

\bibitem{MorFes}
P.~M.~Morse and  H.~Feshbach, 
Methods of Theoretical Physics, Part I,  
McGraw-Hill, New York, 1953. 

\bibitem{Dekker}
H.~Dekker, 
Classical and quantum mechanics of the damped harmonic oscillator, 
Phys. Rep. {\bf 80}, 1 (1981).  See also references therein. 

\bibitem{Razavy}
M.~Razavy, 
\textit{Classical and Quantum Dissipative Systems}, 2nd Edition, 
(World Scientific, Singapore, 2017).  See also references therein. 

\bibitem{FesTik}
H.~Feshbach and  Y.~Tikochinsky, 
Quantization of the damped harmonic oscillator, 
Transact. N.Y. Acad. Sci. Ser. II {\bf 38}, 44 (1977). 
 
\bibitem{CRV}
E.~Celeghini,  M.~Rasetti,  and G.~Vitiello, 
Quantum dissipation, 
Ann. Phys. {\bf 215}, 156 (1992). 

\bibitem{SVW}
Y.~N.~Srivastava,  G.~Vitiello,  and A.~Widom, 
Quantum dissipation and quantum noise, 
Ann. Phys. {\bf 238}, 200 (1995);    
arXiv:hep-th/9502044. 

\bibitem{BanMuk}
R.~Banerjee and  P.~Mukherjee, 
A canonical approach to the quantization of the damped harmonic oscillator, 
J. Phys. A: Math. Gen. {\bf 35}, 5591 (2002);  	
arXiv:quant-ph/0108055. 

\bibitem{BlaJiz}
M.~Blasone and P.~Jizba, 
Bateman's dual system revisited: quantization, geometric phase and relation 
with the ground-state energy of the linear harmonic oscillator, 
Ann. Phys. {\bf 312}, 354 (2004);   
arXiv:quant-ph/0102128. 

\bibitem{ChrJur}
D.~Chru\'{s}ci\'{n}ski and  J.~Jurkowski, 
Quantum damped oscillator I: dissipation and resonances, 
Ann. Phys. {\bf 321}, 854 (2006);   
arXiv:quant-ph/0506007. 

\bibitem{Chruscinski}
D.~Chru\'{s}ci\'{n}ski, 
Quantum damped oscillator II: Bateman's Hamiltonian vs. 2D parabolic potential barrier, 
Ann. Phys. {\bf 321}, 840 (2006);  
arXiv:quant-ph/0506091.  

\bibitem{MajSuz}
H.~Majima and A.~Suzuki, 
Quantization and instability of the damped harmonic oscillator subject to a time-dependent force, 
Ann. Phys. {\bf 326}, 3000 (2011). 

\bibitem{GLAC}
J.~Guerrero,  F.~F.~L\'{o}pez-Ruiz,  V.~Aldaya,  and F.~Coss\'{\i}o, 
Symmetries of the quantum damped harmonic oscillator, 
J. Phys. A: Math. Theor. {\bf 45}, 475303 (2012);  
arXiv:1210.4058 [math-ph]. 

\bibitem{PNC}
S.~K.~Pal,  P.~Nandi, and B.~Chakraborty, 
Connecting dissipation and noncommutativity: A Bateman system case study, 
Phys. Rev. A {\bf 97}, 062110 (2018); 
arXiv:1803.03334 [quant-ph]. 

\bibitem{DegFuj} 
S.~Deguchi, Y.~Fujiwara, and K.~Nakano, 
Two quantization approaches to the Bateman oscillator model, 
Ann. Phys. {\bf 403}, 34 (2019);  
arXiv:1807.04403 [quant-ph]. 

\bibitem{Caldirola}
P.~Caldirola, 
Forze non conservative nella meccanica quantistica, 
Nuovo Cim. {\bf 18}, 393 (1941).

\bibitem{Kanai}
E.~Kanai, 
On the quantization of the dissipative systems, 
Prog. Theor. Phys. {\bf 3}, 440 (1948). 

\bibitem{Kerner} 
E.~H.~Kerner, 
Note on the forced and damped oscillator in quantum mechanics, 
Can. J. Phys. {\bf 36}, 371 (1958).

\bibitem{Hasse}
R.~W.~Hasse, 
On the quantum mechanical treatment of dissipative systems, 
J. Math. Phys. {\bf 16}, 2005 (1975).

\bibitem{Choi}
J.~R.~Choi, 
The Decay Properties of a Single-photon in Linear Media, 
Chinese J. Phys. {\bf 41}, 257 (2003). 

\bibitem{BFG}
M.~C.~Baldiotti, R.~Fresneda, and D.~M.~Gitman, 
Quantization of the damped harmonic oscillator revisited, 
Phys. Lett. A {\bf 375}, 1630 (2011); 
arXiv:1005.4096 [quant-ph].   

\bibitem{Dirac} 
P.~A.~M.~Dirac, \textit{Lectures on Quantum Mechanics} 
(Belfer Graduate School of Science, Yeshiva University, New York 1964). 

\bibitem{HRT} 
A.~J.~Hanson, T.~Regge, and C.~Teitelboim, 
\textit{Constrained Hamiltonian Systems} 
(Accademia Nazionale dei Lincei, Rome, 1976). 

\bibitem{HenTei} 
M.~Henneaux and C.~Teitelboim, 
\textit{Quantization of Gauge Systems} 
(Princeton University Press, Princeton, NJ, 1992). 

\bibitem{Sakurai}
J.~J.~Sakurai, 
\textit{Modern Quantum Mechanics}, Revised Edition 
(Addison-Wesley, Redwood City, CA, 1994). 

\bibitem{OmoKam} 
M.~Omote and S.~Kamefuchi, 
On formal solutions to the Schr\"{o}dinger equation, 
Phys. Lett. A {\bf 206}, 273 (1995).  

\bibitem{KamOmo}
S.~Kamefuchi and M.~Omote, 
\textit{Advanced Quantum Mechanics} 
(Asakura Publishing Co., Ltd., 2003), in Japanese. 
 
\bibitem{Razavy_2}
M.~Razavy,
Quantum-mechanical irreversible motion of an infinite chain,
Can. J. Phys. {\bf 57}, 1731 (1979).

\bibitem{ISSSS}
A.~Isar, A.~Sandulescu, H.~Scutaru, E.~Stefanescu, and W.~Scheid,
Open quantum systems,
Int. J. Mod. Phys. E {\bf 03}, 635 (1994);
arXiv:quant-ph/0411189. 

\bibitem{TakUme}
Y.~Takahashi and  H.~Umezawa, 
Thermo field dynamics, 
Collect. Phenom. {\bf 2}, 55  (1975).

\bibitem{Umezawa}
H.~Umezawa, 
\textit{Advanced Field Theory: Micro, Macro, and Thermal Physics}   
(American Institute of Physics, New York, 1993). 

\bibitem{Galley}
C.~R.~Galley, 
Classical mechanics of nonconservative systems, 
Phys. Rev. Lett. {\bf 110}, 174301 (2013);  
arXiv:1210.2745 [gr-qc]. 

\bibitem{GTS}
C.~R.~Galley, D.~Tsang, L.~C.~Stein, 
The principle of stationary nonconservative action for classical mechanics and field theories, 
arXiv:1412.3082 [math-ph]. 

\bibitem{Polonyi} 
J.~Polonyi, 
Classical and quantum effective theories, 
Phys. Rev. D {\bf 90}, 065010 (2014); 
arXiv:1407.6526 [hep-th]. 

\bibitem{DNS} 
S.~Deguchi, K.~Nakano, and T.~Suzuki, 
Relativistic Lagrangians for the Lorentz-Dirac equation, 
Ann. Phys. {\bf 360}, 539 (2015);  
arXiv:1501.04551 [physics.class-ph]. 

\end{thebibliography}
\end{document}